\documentclass[sigconf]{acmart}

\usepackage{tabularx} 
\usepackage{array,booktabs,multirow} 
\usepackage{graphicx} 
\usepackage{graphics} 

\usepackage{pgfplots} 

\AtBeginDocument{%
  \providecommand\BibTeX{{%
    \normalfont B\kern-0.5em{\scshape i\kern-0.25em b}\kern-0.8em\TeX}}}

\copyrightyear{2023}
\acmYear{2023}
\setcopyright{acmlicensed}
\acmConference[RecSys '23]{Seventeenth ACM Conference on Recommender Systems}{September 18--22, 2023}{Singapore, Singapore}
\acmBooktitle{Seventeenth ACM Conference on Recommender Systems (RecSys '23), September 18--22, 2023, Singapore, Singapore}
\acmPrice{15.00}
\acmDOI{10.1145/3604915.3608859}
\acmISBN{979-8-4007-0241-9/23/09}

\begin{document}

\title{Time-Aware Item Weighting for the Next Basket Recommendations}

\author{Aleksey Romanov}
\orcid{0009-0001-4760-2733}
\affiliation{%
   \institution{National Research University Higher School of Economics}
   \city{Moscow}
   \country{Russia}
}
\email{adromanov_2@edu.hse.ru}

\author{Oleg Lashinin}
\orcid{0000-0001-8894-9592}
\affiliation{%
  \institution{Tinkoff}
  \city{Moscow}
  \country{Russia}}
\email{fotol764@gmail.com}

\author{Marina Ananyeva}
\orcid{0000-0002-9885-2230}
\affiliation{%
  \institution{National Research University Higher School of Economics}
  \city{Moscow}
  \country{Russia}}
\email{ananyeva.me@gmail.com}

\author{Sergey Kolesnikov}
\orcid{0000-0002-4820-987X}
\affiliation{%
  \institution{Tinkoff}
  \city{Moscow}
  \country{Russia}}
\email{s.s.kolesnikov@tinkoff.ru}

\renewcommand{\shortauthors}{A. Romanov, O. Lashinin et al.}

\begin{abstract}
In this paper we study the next basket recommendation problem. Recent methods use different approaches to achieve better performance. However, many of them do not use information about the time of prediction and time intervals between baskets. To fill this gap, we propose a novel method, Time-Aware Item-based Weighting (TAIW), which takes timestamps and intervals into account. We provide experiments on three real-world datasets, and TAIW outperforms well-tuned state-of-the-art baselines for next-basket recommendations. In addition, we show the results of an ablation study and a case study of a few items.
\end{abstract}




\maketitle

\section{Introduction}

Next Basket Recommendation (NBR) has become an important problem in the field of recommender systems due to the growth of e-commerce platforms \cite{linden2003amazon, bhatti2020commerce}. A lot of effort has gone into creating accurate algorithms. However, the current progress in NBR seems to be questionable \cite{li2021next}. User behaviour shows a high degree of repetition across different open source datasets. Items bought in the past tend to appear in the next transactions. This means that we can easily count the personalised frequency of past purchases. These are good predictors for next basket recommendations \cite{li2021next, shao2022systematical, tifuknn, upcf}. However, such a simple heuristic recommends new items in a non-personalised way and does not model the dynamics of the user.



To overcome these and other limitations, researchers are developing different models. According to recent reviews of NBR \cite{shao2022systematical, li2021next}, frequency-based methods outperform many deep learning-based approaches.  A representative example is TIFU-KNN \cite{tifuknn}. This method introduces Personalised Item Frequency (PIF) vectors to represent user interests. The authors introduced two types of weighting interactions to capture the dynamics of user preferences. However, TIFU-KNN has some limitations. The first is that the weights depend only on the ordinal number of the basket. While this is a potential area for improvement, the algorithm does not handle time intervals between baskets. The second limitation is that the weights do not vary between items. If someone bought milk in the last basket, they can buy milk again in the next basket. But if someone has bought washing powder, they may not need it for a long time, because a packet of powder can last a long time. The final possible limitation is that TIFU-KNN does not take into account the time gap between the last known interaction and the time of the prediction. All of these limitations have the potential to limit the performance.

In this paper, we present a novel method called Time-Aware Item-based Weighting (TAIW) for next-basket recommendation that overcomes both limitations. We are inspired by the simplicity and superiority of TIFU-KNN. However, TAIW uses more flexible weights for each item based on the current time of prediction. We use the Hawkes process \cite{hawkes}, which helps to estimate relevance scores for previously purchased items based on time intervals between interactions.  The contributions of this paper can be listed as follows:

\begin{figure*}[t!]
\centering
\includegraphics[width=90mm]{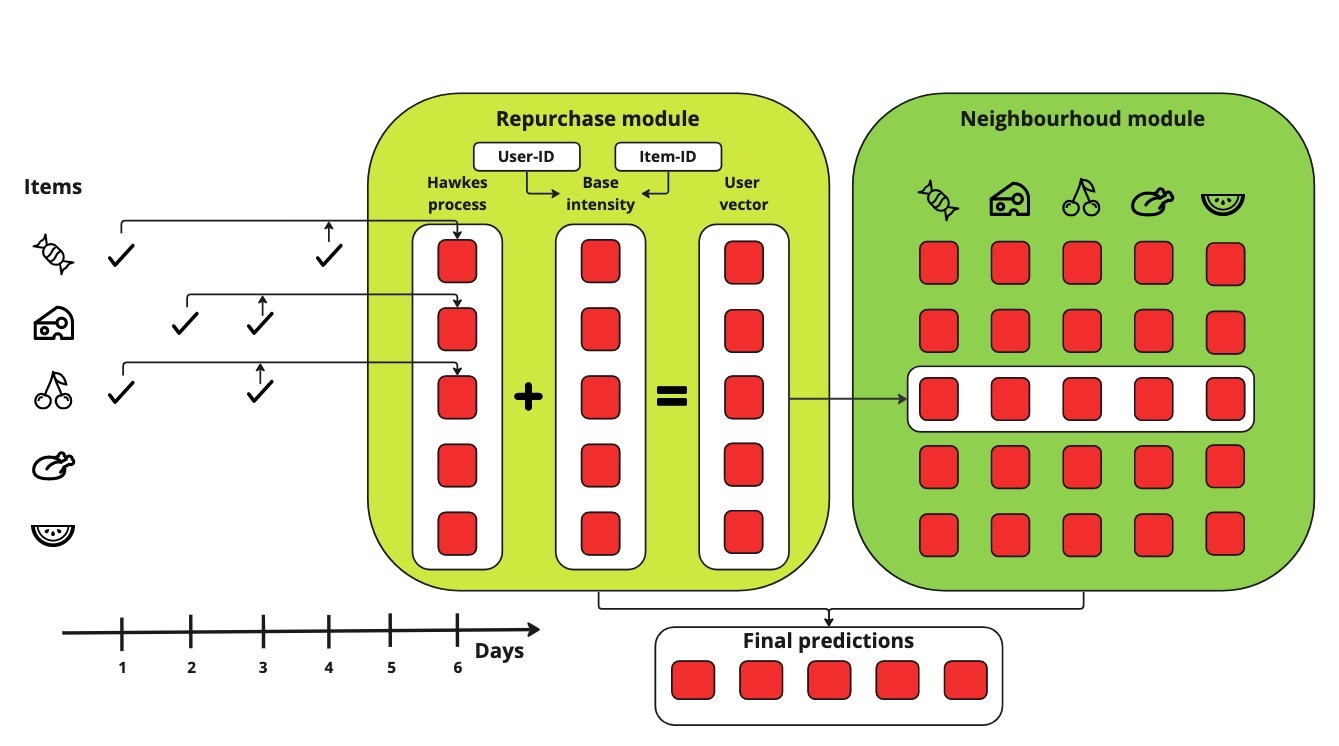}
\caption{A general overview of the proposed TAIW model. It has two modules, the Repurchase Module and the Neighbourhood Module. The Repurchase Module uses the Hawkes process to take into account the contribution of historical purchases of each item at the current moment. Base intensity allows the estimation of relevance scores for both consumed and unconsumed items. The final user vector is the sum of these two vectors. The Neighbourhood Module stores vectors of all users and helps to find preferences of similar users to improve recommendation performance.}
\Description{Fully described in the text.}
\label{fig:taiw_overview}
\end{figure*}

\begin{itemize}
\item We have described a novel method, Time-Aware Item-based Weighting (TAIW for short), for next-basket recommendations. It overcomes the limitations of TIFU-KNN. To encourage reproducibility and future research, we share the implementation of TAIW as well as other baselines in our experiments online.

\item We conducted experiments on three real-world datasets and demonstrated the superiority of TAIW over well-tuned high-performance baselines for the NBR task.

\item We conducted an ablation study, a temporal context importance analysis and a case study to gain insight into TAIW.

\end{itemize}

\section{Related work}
Next basket recommendation is a well-studied problem. Early work used Markov chains \cite{rendle2010factorizing}, recurrent neural networks \cite{yu2016dynamic, recanet} and attention-based mechanisms \cite{repeatnet, hu2019sets2sets}. Such approaches help to consider baskets sequentially in terms of the order in which they appear in user history. Despite their power in other areas of machine learning, recent reviews \cite{li2021next, shao2022systematical} have shown that frequency-based methods still achieve state-of-the-art performance compared to other deep learning methods.


The main reason for this is that high-frequency items are more likely to appear in subsequent baskets due to the repetitive nature of user behaviour. For example, UPCF \cite{upcf} uses purchase frequencies calculated over a fixed time window. This method calculates scores for new items based on some similarity measures such as UserKNN \cite{sarwar2001item} or ItemKNN \cite{wang2006unifying}. Another example of these methods is TIFU-KNN \cite{tifuknn}. The authors introduced Personalised Item Frequency (PIF) vectors and applied the UserKNN approach. PIF vectors have two types of weights based on the ordinal numbers of the baskets. These weights help to capture the dynamics of users' interests. 

However, using only ordinal numbers may limit the quality of recommendations. For example, in the next item recommendation, TiSASRec \cite{li2020time} with explicit modelling of time intervals outperforms SASRec \cite{kang2018self} using only ordinal numbers of interactions. An interesting approach SLRC \cite{slrccite} uses the Hawkes process \cite{hawkes} to estimate the probabilities that interactions will repeat. This allowed the impact of past interactions to be calculated based on the time to the moment of prediction. Similar methods have already been developed for next basket recommendations. ReCaNet \cite{recanet} uses GRU layer to handle time intervals between repurchases of each item. In this work \cite{katz2022learning} a hyper-convolutional model learns purchase cycles and recommends items from history at the right time. Both methods focus only on the repurchase part of the NBR problem, they do not predict new items in the next basket.

Finally, the authors \cite{naumov2023time} introduced a modified TIFU-KNN-TA that works with timestamps and weights interactions based on real time intervals between baskets. However, this model does not learn item-specific weights. In addition, the weights decrease monotonically from more recent to earlier interactions. These two limitations could limit recommendation performance.

\section{Proposed Method}

In this section, we define the NBR problem and describe a new model for this task called Time-Aware Item Specific Weights (TAIW). We are inspired by the aforementioned superiority and simplicity of TIFU-KNN, the time awareness of TIFU-KNN-TA and the item specific weights in SLRC. An overview of TAIW is shown in Figure \ref{fig:taiw_overview}. The proposed method has two modules: the Repurchase Module and the Neighbourhood Module, which are described below. 

\subsection{Problem Definition}

Let $U$ be the set of users and $V$ - the item set. We can represent the consumption history of user $u \in U$ as a sequence $B^{u} = \{(b_{1}^{u}, t_{1}^{u}), ... , (b_{|B^{u}|}^{u}, t_{|B^{u}|}^{u})\}$, where $b_{j}^{u} = \{ v_{1}^{u, j}, ... , v_{|b_{j}^{u}|}^{u, j}\}$ is an unordered set of items (in other words, a basket) purchased by user $u$ at the corresponding timestamp $t_{j}^{u} \in \mathbb {R^{+}}$. We assume that $t_{j}^{u} < t_{j + 1}^{u} \forall j$. Given a target timestamp $t_{|B^{u}| + 1}^{u}$, our goal is to predict the next basket $b_{|B^{u}| + 1}^{u}$ of user $u$.

\subsection{TAIW Overview}

\paragraph{\textup{\textbf{Repurchase Module}}}
As mentioned above, there are many similarities in the repurchase behaviour of different users, namely a short-term and a long-term pattern in the distribution of inter-consumption gaps \cite{slrccite, wang2020toward}. The former refers to the user's desire to repurchase the item immediately after the previous purchase. The latter refers to the lifespan of the purchased item. It is worth noting that these patterns vary from item to item, which means that the target time $t_{|B^{u}| + 1}^{u}$ may be an appropriate time to repurchase some items rather than explore new items. The aim of the Repurchase Module of the proposed method is to learn patterns of inter-consumption gaps for all items and to rank items according to their relevance at the target timestamp $t_{|B^{u}| + 1}^{u}$. To do this, we use a scoring function similar to SLRC \cite{slrccite}, which comes from the Hawkes process \cite{hawkes}. The relevance score $\lambda_{u, i}(t_{|B^{u}| + 1}^{u})$ of item $i \in V$ for user $u \in U$ at the target timestamp $t_{|B^{u}| + 1}^{u}$ can be calculated as follows:

\begin{equation}
    \lambda_{u, i}(t_{|B^{u}| + 1}^{u}) = \lambda_{u, i}^{0} + \alpha_{i} \sum_{(b_{j}^{u}, t_{j}^{u}) \in B^{u}:\ i \in b_{j}^{u},\ t_{j}^{u} < t_{|B^{u}| + 1}^{u}} \gamma_i(t_{|B^{u}| + 1}^{u} - t_{j}^{u}),
\end{equation}
where $\lambda_{u, i}^{0}$ is some base intensity score of user $u$ to buy item $i$. The function $\gamma_i$ is a scoring function of the time interval between the last basket containing item $i$ and the prediction time. It should be proportional to the probability of repurchasing the item after the interval. The parameter $\alpha_{i}$ is an item-specific importance factor of the repurchase component.

To properly rank items at the target time $t_{|B^{u}| + 1}^{u}$ according to their relevance at that time, $\gamma_i$ needs to fit item-specific patterns. Similar to \cite{slrccite}, we model short- and long-term repurchase patterns with the following family of functions:

\begin{equation}\label{gamma}
     \gamma_i(\Delta t) = \pi^i E(\Delta t |\ \beta^i) + (1-\pi^i) \mathcal{N}(\Delta t |\ \mu^i, \sigma^i),
\end{equation}
where $E(\Delta t |\ \beta^i)$ is the probability density function (PDF) of the exponential distribution with parameter $\lambda = \beta^i$, $\mathcal{N}(\Delta t |\ \mu^i, \sigma^i)$ is the PDF of the Gaussian distribution with parameters $\mu = \mu^i$, $\sigma = \sigma^i$, $\pi^i$ is the coefficient of the linear combination of different patterns. According to our research on the datasets considered, for the vast majority of items the interval between purchases is exponentially distributed, indicating the dominance of short-term patterns. However, for some items a high repurchase frequency can be observed after rather long time intervals. As a result, the chosen family of functions could potentially fit the majority of existing pattern forms.

The basic intensity $\lambda_{u, i}^{0}$ of user $u$ to buy item $i$ has been introduced in the SLRC model for recommending new items. The authors used models such as BPR \cite{rendle2012bpr} and NCF  \cite{he2017neural} with trainable user embeddings to model this basic intensity. It can be considered as a limitation in practice \cite{schnabel2022situating} due to the fact that this model cannot be applied to unseen users in an inductive scenario. The proposed method gives the possibility of recommending new items by another module. Therefore, we can consider an inductive version of TAIW without learning user-ID based embeddings:

\begin{equation}
    \lambda_{u, i}(t_{|B^{u}| + 1}^{u}) = \sum_{(b_{j}^{u}, t_{j}^{u}) \in B^{u}:\ i \in b_{j}^{u},\ t_{j}^{u} < t_{|B^{u}| + 1}^{u}} \gamma_i(t_{|B^{u}| + 1}^{u} - t_{j}^{u}).
\end{equation}
We call the corresponding model inductive TAIW or TAIWI. We will show below that TAIWI outperforms other state-of-the-art methods.

\paragraph{\textup{\textbf{Neighbourhood Module}}}
Inspired by the success \cite{li2021next} of user-based k-Nearest Neighbour (kNN) methods (TIFU-KNN \cite{tifuknn} and UPCF \cite{upcf}) for NBR, we introduce a Neighbourhood Module. We define the representation of user $u \in U$ at the target timestamp $t_{|B^{u}| + 1}^{u}$ as follows:

\begin{equation}
    f(u, t_{|B^{u}| + 1}^{u}) = (\lambda_{u, 1}(t_{|B^{u}| + 1}^{u}), ..., \lambda_{u, |V|}(t_{|B^{u}| + 1}^{u}))
\end{equation}
It is then possible to calculate vector representations for other users. Note that neighbours must be calculated without information leakage from test or validation sets. For simplicity, the neighbours' representations can be computed at the timestamps of their last baskets from the training set. As a result, there is a set of other users' representations $\{f(u, t_{|B^{u}|}^{u})|\ u \in U\}$. 

Once we have the target user's representation and the other users' vectors, we can calculate similarity scores between them. Inspired by TIFU-KNN, we define the final prediction vector as a linear combination of the representation of the target user $u$ at the target timestamp $t_{|B^{u}| + 1}^{u}$ and the average vector of the nearest neighbours representations from the set $\{f(u, t_{|B^{u}|}^{u})|\ u \in U\}$:

\begin{equation}
    P_u = \alpha * f(u, t_{|B^{u}| + 1}^{u}) + (1 - \alpha) * \frac{1}{k} \sum_{\hat{u} \in kNN(u)} f(\hat{u}, t_{|B^{\hat{u}}|}^{\hat{u}}),
\end{equation}
where $kNN(u)$ is a set of $k$ users from the training set with the smallest Euclidean distance to the representation of the target user.

\subsection{Training Setup}

The Repurchase Module of the proposed models has a number of trainable parameters. These are $\theta = \{\vec{\pi}, \vec{\beta}, \vec{\mu}, \vec{\sigma}\}$, which refer to item-specific repurchase patterns. Unlike TAIWI, TAIW has $\vec{\alpha}$ and basic intensity model parameters to train. We use the BPR model to calculate basic intensities in TAIW. Pairwise ranking loss is used to train the parameters of the Repurchase Module:

\begin{equation}
    \mathscr{L} = - \sum_{u \in U} \sum_{(b_{j}^{u}, t_{j}^{u}) \in B^{u}} \sum_{i \in b_{j}^{u}} \ln{\sigma{
    (\lambda_{u, i}(t_{j}^{u}) - \lambda_{u, i^-}(t_{j}^{u}))
    }},
\end{equation}
where $i^- \in V \backslash (b_{1}^{u} \cup b_{2}^{u} \cup ... \cup b_{|B^{u}|}^{u})$ is a random negative element. For the TAIW model we include the standard L2-regularisation of the basic intensity model in $\mathscr{L}$. Moreover, the Neighbourhood Module provides our models with several hyperparameters: $k$ and $\alpha$.  In addition, there are such common hyperparameters as \textit{learning rate}, \textit{batch size}. The TAIW model includes hyperparameters of the basic intensity model.

\section{EXPERIMENTS}

We have provided experiments to answer the following research questions:

\begin{itemize}
    \item \textbf{RQ1}: How does the TAIW model perform against well-tuned baselines for the next basket recommendation task on real-world datasets?
    \item \textbf{RQ2}: How does the quality of TAIW and other methods depend on the time gap between the last known basket and the time of prediction?
    \item \textbf{RQ3}: How do different components of the TAIW model affect performance?
    \item \textbf{RQ4}: What insights can be found in the learned parameters of the model?
    \item \textbf{RQ5}: Can specific repurchase preferences of different users be generalised?

\end{itemize}

\begin{table*}[t!]
\caption{Metrics of the proposed models compared to the baselines. The best and the second best models are indicated by boldface and underline respectively. $\blacktriangle \%$ shows the improvement of our models compared to the best baseline.}
\label{tab:results}
\begin{center}
\begin{tabular}{ w{l}{1em} w{l}{6em} | w{c}{8.0em} w{c}{8.0em} w{c}{8.0em}}
\toprule
\multirow{3}{*}{\rotatebox{90}{Dataset}} & \multicolumn{1}{c|}{\multirow{3}{*}{Baseline}} & \multicolumn{3}{c}{Metrics} \\
&  & Precision@10 &  Recall@10 & NDCG@10\\
& & & &  \\
\midrule
\multirow{8}{*}{\rotatebox{90}{TaFeng}}
& GP-Pop & 0.0572 & 0.1308 & 0.1084 \\
& TIFU-KNN & 0.0574 & 0.1360 & 0.1179\\
& DNNTSP & 0.0502 & 0.1306 & 0.1120\\
& SLRC & 0.0621 & 0.1458 & 0.1194 \\
& TIFU-KNN-TA & 0.0615 & 0.1503 & 0.1248\\
& UPCF & 0.0576 & 0.1365 & 0.1154\\
& TAIW & \underline{0.0644} ($\blacktriangle 3.7\%$) & \underline{0.1565} ($\blacktriangle 4.1\%$) & \textbf{0.1267} ($\blacktriangle 1.5\%$)\\
& TAIWI & \textbf{0.0671} ($\blacktriangle 8.1\%$) & \textbf{0.1642} ($\blacktriangle 9.2\%$) & \underline{0.1261} ($\blacktriangle 1.0\%$)\\
\midrule
\multirow{8}{*}{\rotatebox{90}{TaoBao}}
& GP-Pop & 0.0115 & 0.1116 & 0.0741 \\
& TIFU-KNN & 0.0077 & 0.0749 & 0.0524\\
& DNNTSP & 0.0002 & 0.0016 & 0.0012\\
& SLRC & 0.0116 & 0.1118 & 0.0801 \\
& TIFU-KNN-TA & 0.0078 & 0.0763 & 0.0543\\
& UPCF & 0.0085 & 0.0824 & 0.0553\\
& TAIW & \underline{0.0122} ($\blacktriangle 5.2\%$) & \underline{0.1177} ($\blacktriangle 5.3\%$) & \textbf{0.0815} ($\blacktriangle 1.7\%$)\\
& TAIWI & \textbf{0.0123} ($\blacktriangle 6.0\%$) & \textbf{0.1190} ($\blacktriangle 6.4\%$) & \underline{0.0810} ($\blacktriangle 1.1\%$)\\
\midrule
\multirow{8}{*}{\rotatebox{90}{Dunnhumby}}
& GP-Pop & 0.1091 & 0.1577 & 0.1490 \\
& TIFU-KNN & 0.1157 & 0.1653 & 0.1613\\
& DNNTSP & 0.0613 & 0.0929 & 0.0931\\
& SLRC & 0.1192 & 0.1727 & 0.1675 \\
& TIFU-KNN-TA & 0.1162 & 0.1705 & 0.1593\\
& UPCF & 0.1167 & 0.1663 & 0.1600\\
& TAIW & \textbf{0.1214} ($\blacktriangle 1.8\%$) & \textbf{0.1791} ($\blacktriangle 3.7\%$) & \underline{0.1706} ($\blacktriangle 1.9\%$)\\
& TAIWI & \underline{0.1211} ($\blacktriangle 1.6\%$) & \underline{0.1773} ($\blacktriangle 2.7\%$) & \textbf{0.1713} ($\blacktriangle 2.3\%$)\\
\bottomrule
\end{tabular}
\end{center}
\end{table*}

\subsection{Experimental Settings}

\paragraph{\textup{\textbf{Datasets}}}
We use popular real-world datasets to evaluate our method and other state-of-the-art baselines. \textbf{Dunnhumby}\footnote{\url{https://www.kaggle.com/datasets/frtgnn/dunnhumby-the-complete-journey}} contains household level retail transactions over two years from a group of 2,500 households. \textbf{TaFeng}\footnote{\url{https://www.kaggle.com/datasets/chiranjivdas09/ta-feng-grocery-dataset}} contains transaction data from Chinese grocery stores over 4 months. \textbf{TaoBao}\footnote{\url{https://tianchi.aliyun.com/dataset/649}} is provided by Alibaba and contains user behaviour from Taobao. The preprocessing step in our experiments includes filtering out users and items with few associated interactions (the filtering threshold depends on the dataset). In addition, users who made all their transactions within one day or who made too many transactions (compared to other users) are also removed. The detailed steps can be found online at the link below.

\paragraph{\textup{\textbf{Metrics}}}
We use standard metrics to evaluate ranking quality such as Precision@K, Recall@K and NDCG@K. For NDCG@K we use a standard binary relevance function. 

\paragraph{\textup{\textbf{Baselines}}}
Due to space limitations, only the strongest baselines for NBR are included. \textbf{GP-Pop} is a simple heuristic that recommends items from the history ordered by the frequency of the user's purchases. \textbf{TIFU-KNN} \cite{tifuknn} is a state-of-the-art method for NBR based on the frequency-based User-KNN method. \textbf{UPCF} \cite{upcf} works in a similar way. We include \textbf{TIFU-KNN-TA} \cite{naumov2023time} as a recent improvement to TIFU-KNN. \textbf{DNNTSP} \cite{dnntspcite} is a graph-based model that sometimes outperforms TIFU-KNN, according to \cite{li2021next}. Finally, we add \textbf{SLRC} \cite{slrccite}, which has not been used for NBR benchmarks before, but can be easily applied to NBR without any special effort.

\begin{figure*}[ht!]
\includegraphics[width=\linewidth]{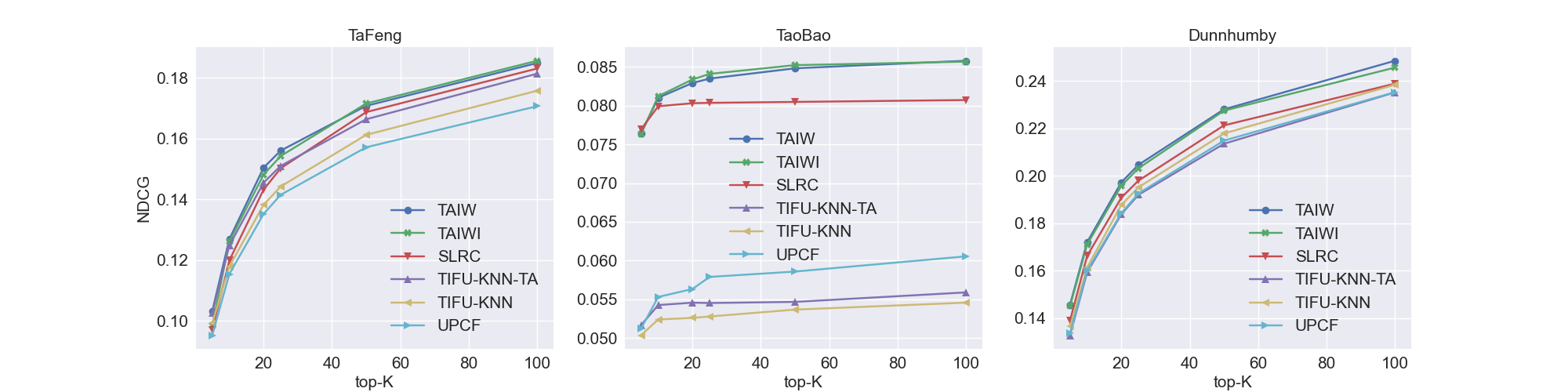}

 \caption{NDCG@K w.r.t. different K values and models across all included datasets.}\label{fig:rq1}
 \Description[TAIW and TAIWI show better performance for different recommendation list sizes]{TAIW and TAIWI outperform the baselines across all datasets for the considered recommendation list sizes from 5 to 100. The superiority increases with the size of the recommendation list.}

\end{figure*}

\begin{figure*}[ht!]
\begin{center}
\includegraphics[width=\linewidth]{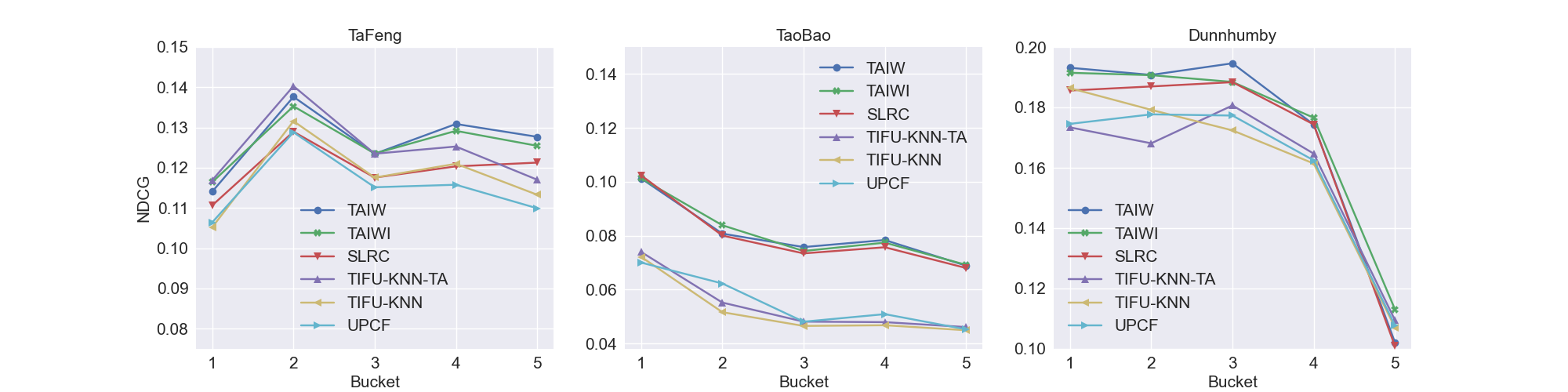}

 \end{center}
 \caption{NDCG@10 w.r.t. different gaps between the last known basket and the target timestamp (all users are divided into 5 equal buckets according to this gap, a larger number of buckets means a larger gap).}\label{fig:rq2}
 \Description{Fully described in the text.}

\end{figure*}

\paragraph{\textup{\textbf{Implementation Details}}}
We looked carefully at the list of frameworks for reproducibility\footnote{\url{https://github.com/ACMRecSys/recsys-evaluation-frameworks}}, but none of them have Next Basket Recommendation baselines. So we tried our best to create reproducible and reliable experiments. We implemented the TAIW and TAIWI models using PyTorch. For the baseline models, we used implementations provided by the authors. The code is available online\footnote{\url{https://github.com/alexeyromanov-hse/time_aware_item_weighting}}. All experiments were conducted using the environment provided by Google Colaboratory\footnote{\url{https://colab.research.google.com/}} (including standard NVIDIA T4 Tensor Core GPUs).

\paragraph{\textup{\textbf{Evaluation protocol}}}
We used a standard leave-one-basket protocol to evaluate the NBR models. We took each user's last basket for testing, the penultimate basket for validation and the remaining baskets for training. The computation of all metrics involved ranking items from $V$ for each test user and recommending the top-k items as the user's next basket. Final metrics are reported for the test dataset using tuned hyperparameters (see below) by repeating the training process with different random seeds and averaging the result (3 - 5 different seeds depending on the dataset).

\paragraph{\textup{\textbf{Hyperparameter tuning}}}
We chose the Optuna framework\footnote{\url{https://optuna.org/}} for the hyperparameter tuning. We followed the same steps for each baseline and the proposed models. Namely, Optuna sampled the same number of different sets of hyperparameters to find the best one (25 for each model). The hyperparameter grid is available online due to space limitations. The best set of hyperparameters is the one that maximises NDCG@10 in the validation dataset.

\subsection{Results}

To answer \textbf{RQ1}, we conducted a series of experiments according to the evaluation protocol described. Table \ref{tab:results} shows the final metrics of the considered baselines and the proposed methods. Our TAIW and TAIWI methods outperform all baselines across all datasets and metrics. In addition, we can define SLRC and TIFU-KNN-TA as the most powerful baseline models. Furthermore, Figure \ref{fig:rq1} shows how the size of the recommendation list affects the metrics for the most powerful baseline models and the proposed methods across all datasets. The experiments conducted show the superiority of the proposed methods over the considered baselines.

\begin{figure*}[ht!]
\begin{center}

\includegraphics[width=\linewidth]{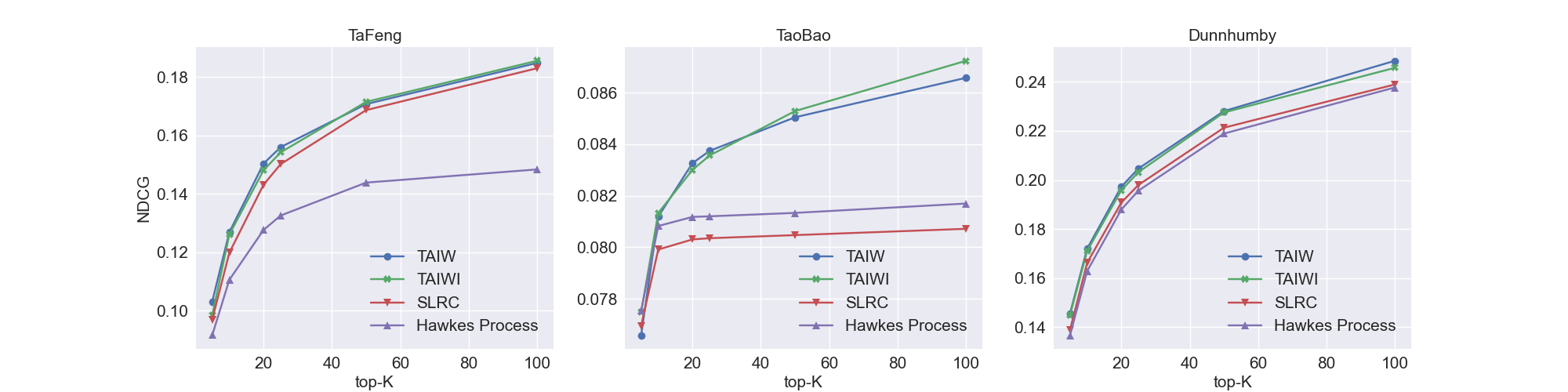}

 \end{center}
 \caption{NDCG@K w.r.t. different K values and configurations across all included datasets.}\label{fig:rq3}
 \Description[Figure shows the importance of the Neighbourhood Module and the small difference between TAIW and TAIWI]{The ablation study shows that the difference between the transductive and inductive variants of TAIW is insignificant and the Neighbourhood Module brings a significant increase in performance for different recommendation list sizes.}

\end{figure*}

\begin{figure*}[ht!]
  \centering
  \includegraphics[width=\linewidth]{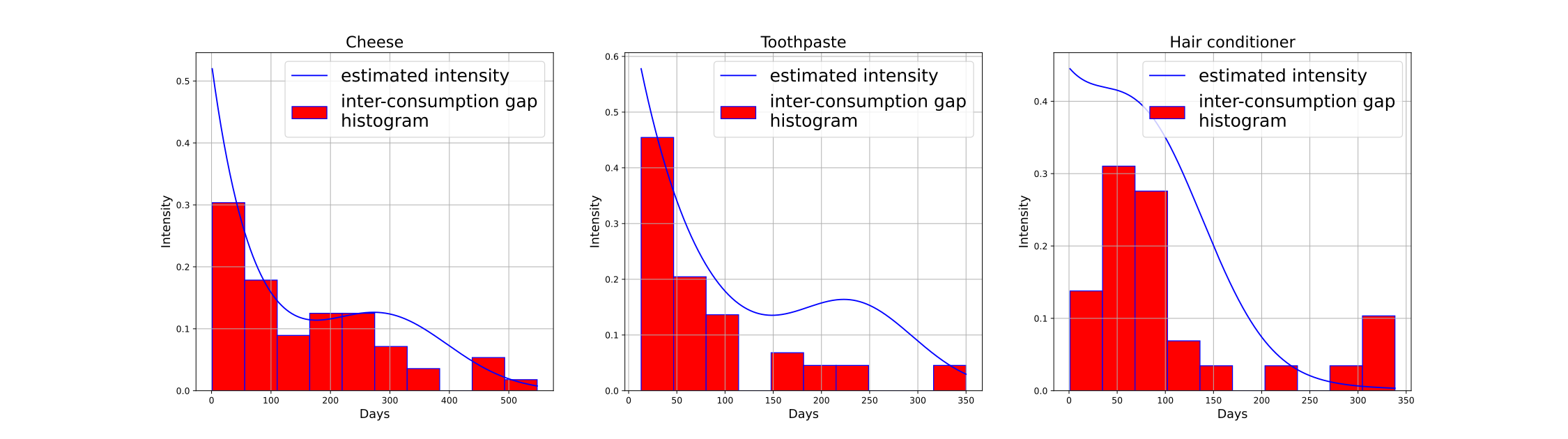}

  \caption{Inter-consumption gap distribution and estimated intensity functions for specific Dunnhumby items.}
  \Description{Fully described in the text.}
  
  \label{fig:items_intensities}
\end{figure*}

To study \textbf{RQ2}, we divided users into five equal-sized buckets according to the number of days between the last training basket and the test basket to examine the predictive power of the considered methods for different time gaps. Figure \ref{fig:rq2} shows the result. User preferences can drift over $t_{|B^{u}| + 1}^{u} - t_{|B^{u}|}^{u}$ due to specific repurchase patterns. As a result, TIFU-KNN may suffer from long gaps. On the other hand, TIFU-KNN-TA, SLRC and TAIW take this gaps into account. We can see that TIFU-KNN-TA is better than TIFU-KNN for longer time intervals. The performance of TAIW and TAIWI degrades more slowly than other models. This may indicate the importance of adjusting the item-specific weights for the temporal context.

\begin{figure*}[ht!]
\begin{center}
\includegraphics[width=\linewidth]{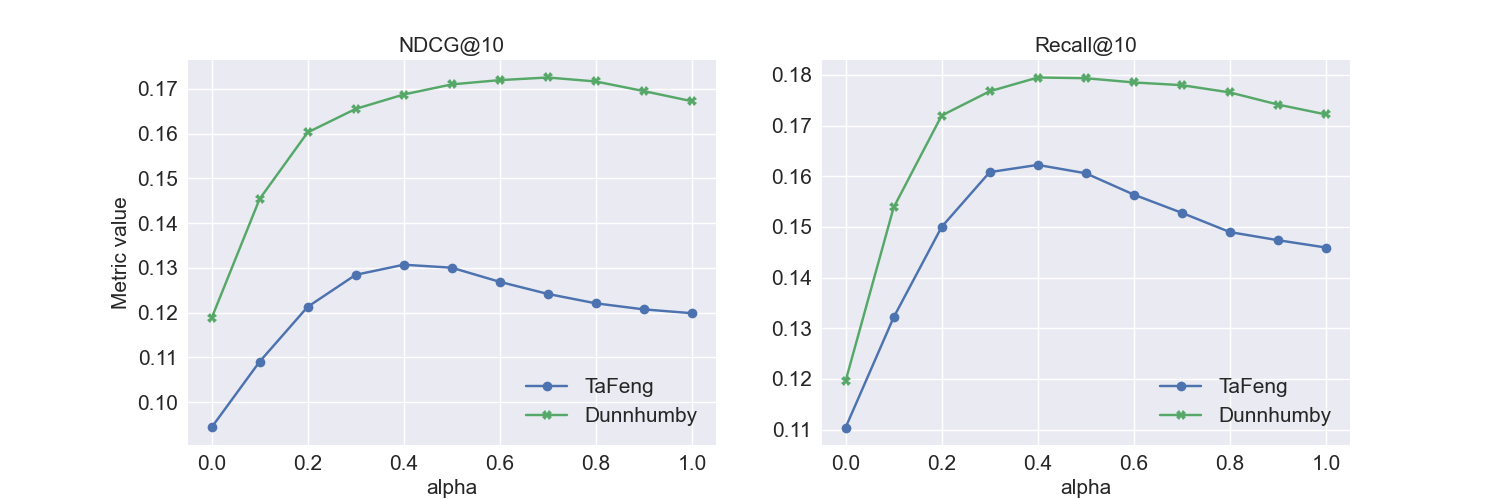}

 \end{center}
 \caption{NDCG@10 and Recall@10 w.r.t. different $\alpha$ hyperparemeter values.}\label{fig:rq5}

\end{figure*}

Figure \ref{fig:rq3} shows the results of an ablation study covering \textbf{RQ3}. A series of experiments were carried out to assess the importance of different components of the proposed models. Firstly, the performance gap between the transductive and inductive variants of the proposed models (TAIW and TAIWI) is investigated. The results demonstrate that the difference between the models is small. As a consequence, the advantages of the inductive model \cite{schnabel2022situating} can be used without loss of predictive power. Secondly, Figure \ref{fig:rq3} shows the performance of both TAIW and TAIWI without the Neighbourhood Module (SLRC and Hawkess respectively). This leads to the conclusion that the use of the Neighbourhood Module brings a significant increase and robustness to the considered metrics across all datasets.

To answer \textbf{RQ4}, we analysed the estimated intensity functions from equation \ref{gamma} for different items from Dunnhumby, as this dataset has interpretable item categories. Figure \ref{fig:items_intensities} shows the results for cheese, toothpaste and hair conditioner. The actual distributions of the consumption gaps and the estimated functions are quite close. We can see that all three items have decreasing short-term repurchase patterns. However, there is an increase in repurchases after 150-250 days for toothpaste and cheese, while hair conditioner is most often repurchased after 50-100 days. It is worth noting that the Dunnhumby dataset includes transactions from different households that are likely to repurchase household goods after a long period of time. In general, TAIW parameters successfully handle different repurchase patterns for each item.

To test \textbf{RQ5}, we analysed the dependence of the predictive power of TAIW on the contribution of the nearest neighbours to the prediction. Figure \ref{fig:rq5} shows how our target metrics (NDCG@10 and Recall@10) change as a function of $\alpha$ hyperparameter of the Neighbourhood Module. The resulting dependence allows us to conclude that repurchase patterns are not fully generalisable. In fact, there is a significant drop in TAIW performance when $\alpha$ takes values close to zero and the predictions depend only on the users' neighbours. This could mean that in practice we have a limited number of different users, which is why searching for very similar neighbours for them can be a challenge. However, the optimal $\alpha$ value is less than 0.5. As a result, the contribution of the neighbourhood is more significant than the contribution of the user representation for the best configuration of TAIW. This could mean that repurchase patterns can still be generalised to some extent.

\section{Conclusion}
In this paper, we propose a novel method TAIW for next basket recommendations with an open source implementation. TAIW addresses the limitations of the current state-of-the-art TIFU-KNN by dealing with timestamps instead of ordinal numbers of baskets. In addition, it uses item-specific weights to predict the relevance scores of items at the time of prediction. According to our experiments with well-tuned state-of-the-art next basket recommenders, TAIW outperforms them by 3\%-8\% on average across three real-world datasets. It shows more stable results when the time gap between the last known training basket and the test basket is large. An ablation study has shown that an inductive version of TAIW performs similarly.

\bibliographystyle{ACM-Reference-Format}
\bibliography{sample-base}


\begin{thebibliography}{24}


\ifx \showCODEN    \undefined \def \showCODEN     #1{\unskip}     \fi
\ifx \showDOI      \undefined \def \showDOI       #1{#1}\fi
\ifx \showISBNx    \undefined \def \showISBNx     #1{\unskip}     \fi
\ifx \showISBNxiii \undefined \def \showISBNxiii  #1{\unskip}     \fi
\ifx \showISSN     \undefined \def \showISSN      #1{\unskip}     \fi
\ifx \showLCCN     \undefined \def \showLCCN      #1{\unskip}     \fi
\ifx \shownote     \undefined \def \shownote      #1{#1}          \fi
\ifx \showarticletitle \undefined \def \showarticletitle #1{#1}   \fi
\ifx \showURL      \undefined \def \showURL       {\relax}        \fi
\providecommand\bibfield[2]{#2}
\providecommand\bibinfo[2]{#2}
\providecommand\natexlab[1]{#1}
\providecommand\showeprint[2][]{arXiv:#2}

\bibitem[Ariannezhad et~al\mbox{.}(2022)]%
        {recanet}
\bibfield{author}{\bibinfo{person}{Mozhdeh Ariannezhad}, \bibinfo{person}{Sami
  Jullien}, \bibinfo{person}{Ming Li}, \bibinfo{person}{Min Fang},
  \bibinfo{person}{Sebastian Schelter}, {and} \bibinfo{person}{Maarten de
  Rijke}.} \bibinfo{year}{2022}\natexlab{}.
\newblock \showarticletitle{ReCANet: A Repeat Consumption-Aware Neural Network
  for Next Basket Recommendation in Grocery Shopping}. In
  \bibinfo{booktitle}{\emph{Proceedings of the 45th International ACM SIGIR
  Conference on Research and Development in Information Retrieval}} (Madrid,
  Spain) \emph{(\bibinfo{series}{SIGIR '22})}. \bibinfo{publisher}{Association
  for Computing Machinery}, \bibinfo{address}{New York, NY, USA},
  \bibinfo{pages}{1240–1250}.
\newblock
\showISBNx{9781450387323}
\urldef\tempurl%
\url{https://doi.org/10.1145/3477495.3531708}
\showDOI{\tempurl}


\bibitem[Bhatti et~al\mbox{.}(2020)]%
        {bhatti2020commerce}
\bibfield{author}{\bibinfo{person}{Anam Bhatti}, \bibinfo{person}{Hamza Akram},
  \bibinfo{person}{Hafiz Basit}, \bibinfo{person}{Ahmed Khan},
  \bibinfo{person}{Syeda Mahwish}, \bibinfo{person}{Raza Naqvi}, {and}
  \bibinfo{person}{Muhammad Bilal}.} \bibinfo{year}{2020}\natexlab{}.
\newblock \showarticletitle{E-commerce trends during COVID-19 Pandemic}.
\newblock \bibinfo{journal}{\emph{International Journal of Future Generation
  Communication and Networking}}  \bibinfo{volume}{13} (\bibinfo{date}{06}
  \bibinfo{year}{2020}).
\newblock


\bibitem[Faggioli et~al\mbox{.}(2020)]%
        {upcf}
\bibfield{author}{\bibinfo{person}{Guglielmo Faggioli}, \bibinfo{person}{Mirko
  Polato}, {and} \bibinfo{person}{Fabio Aiolli}.}
  \bibinfo{year}{2020}\natexlab{}.
\newblock \showarticletitle{Recency Aware Collaborative Filtering for Next
  Basket Recommendation}. In \bibinfo{booktitle}{\emph{Proceedings of the 28th
  ACM Conference on User Modeling, Adaptation and Personalization}} (Genoa,
  Italy) \emph{(\bibinfo{series}{UMAP '20})}. \bibinfo{publisher}{Association
  for Computing Machinery}, \bibinfo{address}{New York, NY, USA},
  \bibinfo{pages}{80–87}.
\newblock
\showISBNx{9781450368612}
\urldef\tempurl%
\url{https://doi.org/10.1145/3340631.3394850}
\showDOI{\tempurl}


\bibitem[He et~al\mbox{.}(2017)]%
        {he2017neural}
\bibfield{author}{\bibinfo{person}{Xiangnan He}, \bibinfo{person}{Lizi Liao},
  \bibinfo{person}{Hanwang Zhang}, \bibinfo{person}{Liqiang Nie},
  \bibinfo{person}{Xia Hu}, {and} \bibinfo{person}{Tat-Seng Chua}.}
  \bibinfo{year}{2017}\natexlab{}.
\newblock \showarticletitle{Neural Collaborative Filtering}. In
  \bibinfo{booktitle}{\emph{Proceedings of the 26th International Conference on
  World Wide Web}} (Perth, Australia) \emph{(\bibinfo{series}{WWW '17})}.
  \bibinfo{publisher}{International World Wide Web Conferences Steering
  Committee}, \bibinfo{address}{Republic and Canton of Geneva, CHE},
  \bibinfo{pages}{173–182}.
\newblock
\showISBNx{9781450349130}
\urldef\tempurl%
\url{https://doi.org/10.1145/3038912.3052569}
\showDOI{\tempurl}


\bibitem[Hu and He(2019)]%
        {hu2019sets2sets}
\bibfield{author}{\bibinfo{person}{Haoji Hu} {and} \bibinfo{person}{Xiangnan
  He}.} \bibinfo{year}{2019}\natexlab{}.
\newblock \showarticletitle{Sets2Sets: Learning from Sequential Sets with
  Neural Networks}. In \bibinfo{booktitle}{\emph{Proceedings of the 25th ACM
  SIGKDD International Conference on Knowledge Discovery \& Data Mining}}
  (Anchorage, AK, USA) \emph{(\bibinfo{series}{KDD '19})}.
  \bibinfo{publisher}{Association for Computing Machinery},
  \bibinfo{address}{New York, NY, USA}, \bibinfo{pages}{1491–1499}.
\newblock
\showISBNx{9781450362016}
\urldef\tempurl%
\url{https://doi.org/10.1145/3292500.3330979}
\showDOI{\tempurl}


\bibitem[Hu et~al\mbox{.}(2020)]%
        {tifuknn}
\bibfield{author}{\bibinfo{person}{Haoji Hu}, \bibinfo{person}{Xiangnan He},
  \bibinfo{person}{Jinyang Gao}, {and} \bibinfo{person}{Zhi-Li Zhang}.}
  \bibinfo{year}{2020}\natexlab{}.
\newblock \showarticletitle{Modeling Personalized Item Frequency Information
  for Next-Basket Recommendation}. In \bibinfo{booktitle}{\emph{Proceedings of
  the 43rd International ACM SIGIR Conference on Research and Development in
  Information Retrieval}} (Virtual Event, China) \emph{(\bibinfo{series}{SIGIR
  '20})}. \bibinfo{publisher}{Association for Computing Machinery},
  \bibinfo{address}{New York, NY, USA}, \bibinfo{pages}{1071–1080}.
\newblock
\showISBNx{9781450380164}
\urldef\tempurl%
\url{https://doi.org/10.1145/3397271.3401066}
\showDOI{\tempurl}


\bibitem[Kang and McAuley(2018)]%
        {kang2018self}
\bibfield{author}{\bibinfo{person}{Wang{-}Cheng Kang} {and}
  \bibinfo{person}{Julian~J. McAuley}.} \bibinfo{year}{2018}\natexlab{}.
\newblock \showarticletitle{Self-Attentive Sequential Recommendation}.
\newblock \bibinfo{journal}{\emph{CoRR}}  \bibinfo{volume}{abs/1808.09781}
  (\bibinfo{year}{2018}).
\newblock
\showeprint[arXiv]{1808.09781}
\urldef\tempurl%
\url{http://arxiv.org/abs/1808.09781}
\showURL{%
\tempurl}


\bibitem[Katz et~al\mbox{.}(2022)]%
        {katz2022learning}
\bibfield{author}{\bibinfo{person}{Ori Katz}, \bibinfo{person}{Oren Barkan},
  \bibinfo{person}{Noam Koenigstein}, {and} \bibinfo{person}{Nir Zabari}.}
  \bibinfo{year}{2022}\natexlab{}.
\newblock \showarticletitle{Learning to Ride a Buy-Cycle: A Hyper-Convolutional
  Model for Next Basket Repurchase Recommendation}. In
  \bibinfo{booktitle}{\emph{Proceedings of the 16th ACM Conference on
  Recommender Systems}} (Seattle, WA, USA) \emph{(\bibinfo{series}{RecSys
  '22})}. \bibinfo{publisher}{Association for Computing Machinery},
  \bibinfo{address}{New York, NY, USA}, \bibinfo{pages}{316–326}.
\newblock
\showISBNx{9781450392785}
\urldef\tempurl%
\url{https://doi.org/10.1145/3523227.3546763}
\showDOI{\tempurl}


\bibitem[Laub et~al\mbox{.}(2015)]%
        {hawkes}
\bibfield{author}{\bibinfo{person}{Patrick~J. Laub}, \bibinfo{person}{Thomas
  Taimre}, {and} \bibinfo{person}{Philip~K. Pollett}.}
  \bibinfo{year}{2015}\natexlab{}.
\newblock \showarticletitle{Hawkes Processes}.
\newblock  (\bibinfo{date}{07} \bibinfo{year}{2015}).
\newblock
\showeprint[arxiv]{1507.02822}~[math.PR]


\bibitem[Li et~al\mbox{.}(2020)]%
        {li2020time}
\bibfield{author}{\bibinfo{person}{Jiacheng Li}, \bibinfo{person}{Yujie Wang},
  {and} \bibinfo{person}{Julian McAuley}.} \bibinfo{year}{2020}\natexlab{}.
\newblock \showarticletitle{Time Interval Aware Self-Attention for Sequential
  Recommendation}. In \bibinfo{booktitle}{\emph{Proceedings of the 13th
  International Conference on Web Search and Data Mining}} (Houston, TX, USA)
  \emph{(\bibinfo{series}{WSDM '20})}. \bibinfo{publisher}{Association for
  Computing Machinery}, \bibinfo{address}{New York, NY, USA},
  \bibinfo{pages}{322–330}.
\newblock
\showISBNx{9781450368223}
\urldef\tempurl%
\url{https://doi.org/10.1145/3336191.3371786}
\showDOI{\tempurl}


\bibitem[Li et~al\mbox{.}(2023)]%
        {li2021next}
\bibfield{author}{\bibinfo{person}{Ming Li}, \bibinfo{person}{Sami Jullien},
  \bibinfo{person}{Mozhdeh Ariannezhad}, {and} \bibinfo{person}{Maarten de
  Rijke}.} \bibinfo{year}{2023}\natexlab{}.
\newblock \showarticletitle{A Next Basket Recommendation Reality Check}.
\newblock \bibinfo{journal}{\emph{ACM Trans. Inf. Syst.}} \bibinfo{volume}{41},
  \bibinfo{number}{4}, Article \bibinfo{articleno}{116} (\bibinfo{date}{apr}
  \bibinfo{year}{2023}), \bibinfo{numpages}{29}~pages.
\newblock
\showISSN{1046-8188}
\urldef\tempurl%
\url{https://doi.org/10.1145/3587153}
\showDOI{\tempurl}


\bibitem[Linden et~al\mbox{.}(2003)]%
        {linden2003amazon}
\bibfield{author}{\bibinfo{person}{Greg Linden}, \bibinfo{person}{Brent Smith},
  {and} \bibinfo{person}{Jeremy York}.} \bibinfo{year}{2003}\natexlab{}.
\newblock \showarticletitle{Amazon.Com Recommendations: Item-to-Item
  Collaborative Filtering}.
\newblock \bibinfo{journal}{\emph{IEEE Internet Computing}}
  \bibinfo{volume}{7}, \bibinfo{number}{1} (\bibinfo{date}{jan}
  \bibinfo{year}{2003}), \bibinfo{pages}{76–80}.
\newblock
\showISSN{1089-7801}
\urldef\tempurl%
\url{https://doi.org/10.1109/MIC.2003.1167344}
\showDOI{\tempurl}


\bibitem[Naumov et~al\mbox{.}(2023)]%
        {naumov2023time}
\bibfield{author}{\bibinfo{person}{Sergey Naumov}, \bibinfo{person}{Marina
  Ananyeva}, \bibinfo{person}{Oleg Lashinin}, \bibinfo{person}{Sergey
  Kolesnikov}, {and} \bibinfo{person}{Dmitry~I. Ignatov}.}
  \bibinfo{year}{2023}\natexlab{}.
\newblock \showarticletitle{Time-Dependent Next-Basket Recommendations}. In
  \bibinfo{booktitle}{\emph{Advances in Information Retrieval: 45th European
  Conference on Information Retrieval, ECIR 2023, Dublin, Ireland, April 2–6,
  2023, Proceedings, Part II}} (Dublin, Ireland).
  \bibinfo{publisher}{Springer-Verlag}, \bibinfo{address}{Berlin, Heidelberg},
  \bibinfo{pages}{502–511}.
\newblock
\showISBNx{978-3-031-28237-9}
\urldef\tempurl%
\url{https://doi.org/10.1007/978-3-031-28238-6_41}
\showDOI{\tempurl}


\bibitem[Ren et~al\mbox{.}(2019)]%
        {repeatnet}
\bibfield{author}{\bibinfo{person}{Pengjie Ren}, \bibinfo{person}{Zhumin Chen},
  \bibinfo{person}{Jing Li}, \bibinfo{person}{Zhaochun Ren},
  \bibinfo{person}{Jun Ma}, {and} \bibinfo{person}{Maarten de Rijke}.}
  \bibinfo{year}{2019}\natexlab{}.
\newblock \showarticletitle{RepeatNet: A Repeat Aware Neural Recommendation
  Machine for Session-Based Recommendation}. In
  \bibinfo{booktitle}{\emph{Proceedings of the Thirty-Third AAAI Conference on
  Artificial Intelligence and Thirty-First Innovative Applications of
  Artificial Intelligence Conference and Ninth AAAI Symposium on Educational
  Advances in Artificial Intelligence}} (Honolulu, Hawaii, USA)
  \emph{(\bibinfo{series}{AAAI'19/IAAI'19/EAAI'19})}. \bibinfo{publisher}{AAAI
  Press}, Article \bibinfo{articleno}{590}, \bibinfo{numpages}{8}~pages.
\newblock
\showISBNx{978-1-57735-809-1}
\urldef\tempurl%
\url{https://doi.org/10.1609/aaai.v33i01.33014806}
\showDOI{\tempurl}


\bibitem[Rendle et~al\mbox{.}(2009)]%
        {rendle2012bpr}
\bibfield{author}{\bibinfo{person}{Steffen Rendle}, \bibinfo{person}{Christoph
  Freudenthaler}, \bibinfo{person}{Zeno Gantner}, {and} \bibinfo{person}{Lars
  Schmidt-Thieme}.} \bibinfo{year}{2009}\natexlab{}.
\newblock \showarticletitle{BPR: Bayesian Personalized Ranking from Implicit
  Feedback}. In \bibinfo{booktitle}{\emph{Proceedings of the Twenty-Fifth
  Conference on Uncertainty in Artificial Intelligence}} (Montreal, Quebec,
  Canada) \emph{(\bibinfo{series}{UAI '09})}. \bibinfo{publisher}{AUAI Press},
  \bibinfo{address}{Arlington, Virginia, USA}, \bibinfo{pages}{452–461}.
\newblock
\showISBNx{9780974903958}


\bibitem[Rendle et~al\mbox{.}(2010)]%
        {rendle2010factorizing}
\bibfield{author}{\bibinfo{person}{Steffen Rendle}, \bibinfo{person}{Christoph
  Freudenthaler}, {and} \bibinfo{person}{Lars Schmidt-Thieme}.}
  \bibinfo{year}{2010}\natexlab{}.
\newblock \showarticletitle{Factorizing Personalized Markov Chains for
  Next-Basket Recommendation}. In \bibinfo{booktitle}{\emph{Proceedings of the
  19th International Conference on World Wide Web}} (Raleigh, North Carolina,
  USA) \emph{(\bibinfo{series}{WWW '10})}. \bibinfo{publisher}{Association for
  Computing Machinery}, \bibinfo{address}{New York, NY, USA},
  \bibinfo{pages}{811–820}.
\newblock
\showISBNx{9781605587998}
\urldef\tempurl%
\url{https://doi.org/10.1145/1772690.1772773}
\showDOI{\tempurl}


\bibitem[Sarwar et~al\mbox{.}(2001)]%
        {sarwar2001item}
\bibfield{author}{\bibinfo{person}{Badrul Sarwar}, \bibinfo{person}{George
  Karypis}, \bibinfo{person}{Joseph Konstan}, {and} \bibinfo{person}{John
  Riedl}.} \bibinfo{year}{2001}\natexlab{}.
\newblock \showarticletitle{Item-Based Collaborative Filtering Recommendation
  Algorithms}. In \bibinfo{booktitle}{\emph{Proceedings of the 10th
  International Conference on World Wide Web}} (Hong Kong, Hong Kong)
  \emph{(\bibinfo{series}{WWW '01})}. \bibinfo{publisher}{Association for
  Computing Machinery}, \bibinfo{address}{New York, NY, USA},
  \bibinfo{pages}{285–295}.
\newblock
\showISBNx{1581133480}
\urldef\tempurl%
\url{https://doi.org/10.1145/371920.372071}
\showDOI{\tempurl}


\bibitem[Schnabel et~al\mbox{.}(2022)]%
        {schnabel2022situating}
\bibfield{author}{\bibinfo{person}{Tobias Schnabel}, \bibinfo{person}{Mengting
  Wan}, {and} \bibinfo{person}{Longqi Yang}.} \bibinfo{year}{2022}\natexlab{}.
\newblock \showarticletitle{Situating Recommender Systems in Practice: Towards
  Inductive Learning and Incremental Updates}.
\newblock  (\bibinfo{year}{2022}).
\newblock
\showeprint[arxiv]{2211.06365}~[cs.IR]


\bibitem[Shao et~al\mbox{.}(2022)]%
        {shao2022systematical}
\bibfield{author}{\bibinfo{person}{Zhufeng Shao}, \bibinfo{person}{Shoujin
  Wang}, \bibinfo{person}{Qian Zhang}, \bibinfo{person}{Wenpeng Lu},
  \bibinfo{person}{Zhao Li}, {and} \bibinfo{person}{Xueping Peng}.}
  \bibinfo{year}{2022}\natexlab{}.
\newblock \showarticletitle{A Systematical Evaluation for Next-Basket
  Recommendation Algorithms}. In \bibinfo{booktitle}{\emph{2022 IEEE 9th
  International Conference on Data Science and Advanced Analytics DSAA'2022}},
  \bibfield{editor}{\bibinfo{person}{{Joshua Zhexue} Huang},
  \bibinfo{person}{Yi~Pan}, \bibinfo{person}{Barbara Hammer},
  \bibinfo{person}{{Muhammad Khurram} Khan}, \bibinfo{person}{Xing Xie},
  \bibinfo{person}{Laizhong Cui}, {and} \bibinfo{person}{Yulin He}} (Eds.).
  \bibinfo{publisher}{Institute of Electrical and Electronics Engineers
  (IEEE)}, \bibinfo{address}{United States}, \bibinfo{pages}{1041--1050}.
\newblock
\showISBNx{9781665473316}
\urldef\tempurl%
\url{https://doi.org/10.1109/DSAA54385.2022.10032359}
\showDOI{\tempurl}
\newblock
\shownote{9th IEEE International Conference on Data Science and Advanced
  Analytics, DSAA 2022 ; Conference date: 13-10-2022 Through 16-10-2022}.


\bibitem[Wang et~al\mbox{.}(2021)]%
        {wang2020toward}
\bibfield{author}{\bibinfo{person}{Chenyang Wang}, \bibinfo{person}{Weizhi Ma},
  \bibinfo{person}{Min Zhang}, \bibinfo{person}{Chong Chen},
  \bibinfo{person}{Yiqun Liu}, {and} \bibinfo{person}{Shaoping Ma}.}
  \bibinfo{year}{2021}\natexlab{}.
\newblock \showarticletitle{Toward Dynamic User Intention: Temporal
  Evolutionary Effects of Item Relations in Sequential Recommendation}.
\newblock \bibinfo{journal}{\emph{ACM Transactions on Information Systems
  (TOIS)}} \bibinfo{volume}{39}, \bibinfo{number}{2}, Article
  \bibinfo{articleno}{16} (\bibinfo{date}{dec} \bibinfo{year}{2021}),
  \bibinfo{numpages}{33}~pages.
\newblock
\showISSN{1046-8188}
\urldef\tempurl%
\url{https://doi.org/10.1145/3432244}
\showDOI{\tempurl}


\bibitem[Wang et~al\mbox{.}(2019)]%
        {slrccite}
\bibfield{author}{\bibinfo{person}{Chenyang Wang}, \bibinfo{person}{Min Zhang},
  \bibinfo{person}{Weizhi Ma}, \bibinfo{person}{Yiqun Liu}, {and}
  \bibinfo{person}{Shaoping Ma}.} \bibinfo{year}{2019}\natexlab{}.
\newblock \showarticletitle{Modeling Item-Specific Temporal Dynamics of Repeat
  Consumption for Recommender Systems}. In \bibinfo{booktitle}{\emph{The World
  Wide Web Conference}} (San Francisco, CA, USA) \emph{(\bibinfo{series}{WWW
  '19})}. \bibinfo{publisher}{Association for Computing Machinery},
  \bibinfo{address}{New York, NY, USA}, \bibinfo{pages}{1977–1987}.
\newblock
\showISBNx{9781450366748}
\urldef\tempurl%
\url{https://doi.org/10.1145/3308558.3313594}
\showDOI{\tempurl}


\bibitem[Wang et~al\mbox{.}(2006)]%
        {wang2006unifying}
\bibfield{author}{\bibinfo{person}{Jun Wang}, \bibinfo{person}{Arjen~P. de
  Vries}, {and} \bibinfo{person}{Marcel J.~T. Reinders}.}
  \bibinfo{year}{2006}\natexlab{}.
\newblock \showarticletitle{Unifying User-Based and Item-Based Collaborative
  Filtering Approaches by Similarity Fusion}. In
  \bibinfo{booktitle}{\emph{Proceedings of the 29th Annual International ACM
  SIGIR Conference on Research and Development in Information Retrieval}}
  (Seattle, Washington, USA) \emph{(\bibinfo{series}{SIGIR '06})}.
  \bibinfo{publisher}{Association for Computing Machinery},
  \bibinfo{address}{New York, NY, USA}, \bibinfo{pages}{501–508}.
\newblock
\showISBNx{1595933697}
\urldef\tempurl%
\url{https://doi.org/10.1145/1148170.1148257}
\showDOI{\tempurl}


\bibitem[Yu et~al\mbox{.}(2016)]%
        {yu2016dynamic}
\bibfield{author}{\bibinfo{person}{Feng Yu}, \bibinfo{person}{Qiang Liu},
  \bibinfo{person}{Shu Wu}, \bibinfo{person}{Liang Wang}, {and}
  \bibinfo{person}{Tieniu Tan}.} \bibinfo{year}{2016}\natexlab{}.
\newblock \showarticletitle{A Dynamic Recurrent Model for Next Basket
  Recommendation}. In \bibinfo{booktitle}{\emph{Proceedings of the 39th
  International ACM SIGIR Conference on Research and Development in Information
  Retrieval}} (Pisa, Italy) \emph{(\bibinfo{series}{SIGIR '16})}.
  \bibinfo{publisher}{Association for Computing Machinery},
  \bibinfo{address}{New York, NY, USA}, \bibinfo{pages}{729–732}.
\newblock
\showISBNx{9781450340694}
\urldef\tempurl%
\url{https://doi.org/10.1145/2911451.2914683}
\showDOI{\tempurl}


\bibitem[Yu et~al\mbox{.}(2020)]%
        {dnntspcite}
\bibfield{author}{\bibinfo{person}{Le Yu}, \bibinfo{person}{Leilei Sun},
  \bibinfo{person}{Bowen Du}, \bibinfo{person}{Chuanren Liu},
  \bibinfo{person}{Hui Xiong}, {and} \bibinfo{person}{Weifeng Lv}.}
  \bibinfo{year}{2020}\natexlab{}.
\newblock \showarticletitle{Predicting Temporal Sets with Deep Neural
  Networks}. In \bibinfo{booktitle}{\emph{Proceedings of the 26th ACM SIGKDD
  International Conference on Knowledge Discovery \& Data Mining}} (Virtual
  Event, CA, USA) \emph{(\bibinfo{series}{KDD '20})}.
  \bibinfo{publisher}{Association for Computing Machinery},
  \bibinfo{address}{New York, NY, USA}, \bibinfo{pages}{1083–1091}.
\newblock
\showISBNx{9781450379984}
\urldef\tempurl%
\url{https://doi.org/10.1145/3394486.3403152}
\showDOI{\tempurl}


\end{thebibliography}


\end{document}